\documentstyle[aps]{revtex}

\def\be {\begin{equation}}
\def\ee {\end{equation}}
\def\ba {\begin{eqnarray}}
\def\ea {\end{eqnarray}}
\def\nn {\nonumber}
\def\a  {\alpha}
\def\b  {\beta}
\def\c  {\gamma}

\def\L  {\Lambda}

\def\O  {\Omega}
\def\p  {\pi}

\def\r  {\rho}
\def\th {\theta}

\def\la {\label}
\def\le {\left}
\def\ri {\right}
\def\pa {\partial}
\def\f {\frac}

\def\no {\noindent}
\def\bi {\begin{itemize}}
\def\ei {\end{itemize}}

\def\vs {\vspace}

\begin{document}
\draft
\title{Leading Log Corrections to Bekenstein-Hawking Entropy}
\author{ Saurya Das }
\address{ Dept. of Mathematics and Statistics, } 
\address{University of New Brunswick,}  
\address{Fredericton, New Brunswick-
E3B 5A3, CANADA }
\address{ EMail: saurya@math.unb.ca   }

\maketitle
\thispagestyle{empty}

\begin{abstract}

We show that the Bekenstein-Hawking entropy associated with any black hole
undergoes logarithmic corrections when small thermodynamic fluctuations around 
equilibrium are taken into account. Thus, the corrected expression for black hole 
entropy is given by $S= A/4 - k \ln(A)$, where $A$ is the horizon area and $k$ is 
a constant which depends on the specific black hole. We apply our result to 
BTZ black hole, for which $k=3/2$, as found earlier, as well as to anti-de Sitter-
Schwarzschild and Reissner-Nordstrom black hole in arbitrary spacetime dimensions.
Finally, we examine the role of conformal field theory in black hole entropy and
its corrections. 

\end{abstract}


\section{Introduction}

It is well known that for large black holes, the associated Bekenstein-Hawking
entropy is proportional to its horizon area \cite{bh,string,ash1,car2,solo}. There has been 
attempts to calculate the leading order corrections to this entropy as one
reduces the size of the black hole, using different approaches.
Remarkably, in each case this correction was found to be of the 
form $-k \ln(A)$ 
\cite{qg,qg1,fursaev,carlip,gks,other,solo1,jy,bss,k}. 
Is there an underlying reason for this generic logarithmic correction?
It was shown in \cite{dmb}
that logarithmic corrections to thermodynamic
entropy arise in {\it all} thermodynamic systems when small stable fluctuations
around equilibrium are taken into account. 
The stability condition is equivalent to the specific
heat being positive, so that the corresponding canonical ensemble is stable.
When applied to black holes, 
they indeed predict the form $-k\ln (A)$.
Also, it was shown in \cite{dmb} as to
how these log corrections (including correct coefficients)  can result
from a microscopic theory of quantum gravity. 
In this talk, we will summarize the above results.  
Note that we
still consider the black holes to be large compared to the Planck scale. 
This ensures that the logarithmic terms are indeed much smaller compared
to the Bekenstein-Hawking term.

Let us begin by considering a canonical ensemble with partition function
\be
Z(\b)
= \int_0^\infty \r(E) e^{-\b E} dE~~,
\label{part1}
\ee
where $T=1/\b$ is the temperature in units of the Boltzmann constant
$k_B$.    
The density of states can be obtained from (\ref{part1})
by doing an inverse Laplace transform (keeping $E$ fixed) 
\cite{bm,rkb}:
\be
\r(E) = \f{1}{2\p i} \int _{c-i\infty}^{c+i\infty} Z(\b) e^{\b E} d\b
= \f{1}{2\p i} \int_{c-i\infty}^{c+i\infty} e^{S(\b)}d\b~~,
\la{density1}
\ee
where
\be
S(\b) = \ln Z(\b) + \b E
\label{basic}
\ee
is the {\it exact} entropy
as a function of temperature, not just its value at equilibrium 
(which is the sum of entropies of subsystems of
the thermodynamical system, which are themselves in internal
equilibrium). The complex integral can be performed by the method of steepest
descent around the saddle point $\b_0 (= 1/T_0, T_0$ being the equilibrium 
temperature), such that
$S'_0 := (\pa S(\b)/\pa \b)_{\b=\b_0}=0$. 
Note that the usual equilibrium relation
$E=-(\pa \ln Z(\b)/\pa \b)_{\b=\b_0}$ is obeyed. Expanding $S(\b)$ about
$\b=\b_0$, we get
\be
S = S_0 + \f{1}{2} (\b - \b_0) ^2 S_0'' + \cdots ~~,
\label{ent1}
\ee
where $S_0 := S(\beta_0)$ and $S''_0 := (\pa^2 S(\b)/\pa \b^2)_{\b=\b_0}$.
Substituting (\ref{ent1}) in (\ref{density1}) :
\be
\r(E) = \f{e^{S_0}}{2\p i} \int_{c-i\infty}^{c+i\infty}
e^{1/2 (\b - \b_0)^2 S_0''} d\b ~.
\ee
By choosing $c=\beta_0$, and putting $(\beta-\beta_0)= i x$, where $x$ is
a real variable, we obtain, for $S''_0>0$,
\be
\r(E) = \f{e^{S_0}}{\sqrt{2\p S''_0}} ~~.
\la{corr0}
\ee
Note that the density of states $\r(E)$ and $S_0''$ have dimensions of inverse
energy and energy squared respectively. Henceforth, 
we set the Boltzmann constant $k_B$ to unity.
The logarithm of the density of states $\r(E)$ is then the microcanonical
entropy 
\be
{\cal S} := \ln \r(E) = S_0 - \f{1}{2} \ln S_0'' + ~\mbox{(higher order terms)}.
\la{corr1}
\ee
Here ${\cal S}$ is the corrected microcanonical entropy {\it at equilibrium},
obtained by incorporating small fluctuations around thermal equilibrium.
This is different from the function $S(\b)$, which is the
entropy at any temperature (and not just at equilibrium). When applied to
black holes,  $S_0$ is identified with the Bekenstein-Hawking area law. It is seen that the
correction solely depends on the quantity $S_0''$. Moreover, the second
term in Eq.(\ref{corr1}) appears
with a negative sign, as conjectured in general in \cite{qg1} from
the holographic point of view. Now, we will
estimate $S_0''$, without assuming any specific form of $S(\b)$.
From Eq.(\ref{basic}), it follows that
\be
S''(\b)={1\over Z}({\pa^2 Z(\b)\over {\pa \b^2}})-
{1\over Z^2}({\pa Z\over \pa \b})^2~.
\ee
This means that $S_0''$ is nothing but the fluctuation squared of energy
from the equilibrium, i.e., 
\be
S_0''= <E^2>-<E>^2~, 
\ee
where, by the definition
of $\b_0$, $E=<E>= - (\pa \ln Z/\pa \b)_{\b=\b_0}$. It immediately follows
that
\be 
S_0''= T^2 C
\label{esci}
\ee
where $C \equiv (\partial E/ \partial T)_{T_0}$ is the dimensionless
specific heat. The above result also follows from a detailed 
fluctuation analysis of a stable thermodynamic system \cite{ll}.
 
Substituting for $S_0''$ from (\ref{esci}) in (\ref{corr1}), we get:
\be
{\cal S} = \ln \r = S_0 - \f{1}{2} \ln \le( C~T^2 \ri) + \cdots
\la{corr3}
\ee

We will now apply this formula to black holes.  
For dimensional correctness, it is understood
that the quantity within the logarithm is divided by $k_B^2$.
Naturally, we would replace $T \rightarrow  T_H$, the Hawking temperature. 
Then we calculate $C$ for the specific black holes. It can be shown that 
(\ref{corr3})  is valid so long as the Hawking temperature exceeds
inverse Planck length $L_{Pl}$  \cite{dmb} :
\be
 T >> \f{1}{L_{Pl}}~~.
\la{limit}
\ee
Thus, it is no longer valid near the extremal limit. This is expected since
near extremal black holes are highly quantum objects \cite{ddr,bdk}.  

\section{Applications}

\subsection{BTZ Black Hole}

First let us consider a non-rotating  
BTZ black hole in $3$-dimensions with metric \cite{btz}:
\be
ds^2 = - \le( \f{r^2}{\ell^2} - 8G_3 M \ri) dt^2
+ \le( \f{r^2}{\ell^2} - 8G_3 M \ri)^{-1}  dr^2
+ r^2 d\th^2 ~~.
\ee
Its Bekenstein-Hawking
entropy and Hawking temperature are given by 
\ba
S_0 &=& \f{2\p r_+ }{4 G_3}  \la{btz3} \\
T_H &=& \f{r_+}{2\p \ell^2} = \le[ \f{G_3}{\p^2\ell^2}\ri] ~S_0
\la{temp3}
\ea
where $r_+ = \sqrt{8G_3 M} \ell$ is horizon radius 
($G_3=3$-dimensional Newton's constant), $M$ being mass of the 
black hole and $\ell$ is related to the cosmological constant by: 
$\L =-1/\ell^2$. 
The first law in this case is:
\be
dM = T_H dS_0 ~ .
\label{firstlaw} 
\ee
Thus, the specific heat is :
\be
C = \f{dM}{dT_H} = S_0 = \le[ \f{\p^2 \ell^2 }{G_3} \ri]~T_H 
\la{temp4}
\ee
Plugging in (\ref{temp3}) and (\ref{temp4}) in (\ref{corr3}), we get. 
\ba
{\cal S} = \ln \r &=& S_0 - \f12 \ln \le(S_0  S_0^2\ri) + \cdots \nn \\
&=& S_0 - \f{3}{2} \ln S_0 + \cdots~ \la{btz4}.
\ea
Thus, the correction term is proportional to the 
logarithm of horizon area with proportionality constant 
$-3/2$ is exactly that found in \cite{carlip}.
Introduction of angular momentum is straightforward, and it can be shown that
for rotating BTZ black holes far away from extremality,  the entropy is \cite{dmb}: 
\be
{\cal S} = \ln\r = S_0 - \f{3}{2}\ln S_0 
- \f{(J^2/M^{3/2})}{2S_0} + \cdots ~~~.
\ee

\subsection{AdS-Schwarzschild Black Hole}

Since Schwarzschild black holes have negative specific heat, 
they are unstable against thermal fluctuations \cite{euclidean,hp}. 
Thus (\ref{corr3}) is inapplicable. Instead, we will consider its close cousin: 
anti-de Sitter-Schwarzschild black holes. 

We parametrize the cosmological constant in $d$-dimensions in terms of
the length scale $\ell$ as: 
$\L = -(d-1)(d-2)/2\ell^2$. The corresponding Einstein's equations admit
of black holes solutions 
which are asymptotically anti-de Sitter and whose metric is given by:
\be
ds^2 = - 
\le( 1 - \f{16\p G_d M}{(d-2)\O_{d-2}r^{d-3}} + \f{r^2}{\ell^2} \ri) dt^2
+ \le( 1 - \f{16\p G_d M}{(d-2)\O_{d-2}r^{d-3}} + \f{r^2}{\ell^2} \ri)^{-1} 
dr^2 + r^2 d\O_{d-2}^2 ~~.
\ee
Here the parameter $M$ is the conserved charge associated with the 
timelike Killing vector (`mass') \cite{ash2}. The entropy and Hawking 
temperature are given by:
\ba
S_0 &=& \f{\O_{d-2} r_+^{d-2}}{4G_d} \\
T_H &=& \f{(d-1) r_+^2 + (d-3)\ell^2}{4\p \ell^2 r_+}
= \le[ \f{d-3}{4\p} \le( \f{\O_{d-2}}{4G_d} \ri)^{1/(d-2)} \ri]~
\le[ 1 + \f{(d-1) r_+^2}{(d-3) \ell^2}     \ri]~
S_0^{-1/(d-2)} ~~.
\la{adstemp1}
\ea
Although the function $r_+=r_+(M)$ is not explicit,
$C$ can still be calculated using : 
$$ C = \f{(d-2) \O_{d-2} r_+^{d-3}}{4 G_d} 
\le[ \f{\pa \ln T_H }{ \pa r_+}  \ri] ~~~,$$
which yields \cite{mann} :
\be
C =  \le[  (d-2) \f{(d-1)r_+^2/\ell^2 + (d-3) }{ (d-1)r_+^2/\ell^2 
- (d-3) } \ri]~S_0 ~~.
\la{adscv1}
\ee
Note that the limit $\ell \rightarrow \infty$ reproduces the 
(negative) Schwarzschild value of $C$. However, for
\be
 \ell < {r_+}{\sqrt{\f{d-1}{d-3}}}~~~, 
\la{ineq1}
\ee
$C$ changes signature and becomes
positive. In the limit of large $\L$ ($\ell \rightarrow 0$), 
$C$ approaches a positive finite value:
\be
C  = (d-2) S_0~~. 
\la{adscv2}
\ee
Consequently, the canonical ensemble becomes stable against 
small fluctuations around equilibrium \cite{hp}. In other
words, introduction of a negative cosmological constant remedies the otherwise 
unstable Schwarzschild canonical ensemble. 
Let us calculate the leading order corrections in this limit, although
one can do so for any $\ell$ satisfying (\ref{ineq1}). From 
(\ref{corr3}), (\ref{adstemp1}) and (\ref{adscv2}), we get:
\be
{\cal S} = S_0 - \f{d}{2(d-2)} \ln S_0  + \cdots  
\la{scadscorr2}
\ee
Thus, once again the leading corrections are logarithmic. 
It is clear that logarithmic corrections result for all values
of $\ell$ lying in the range (\ref{ineq1}). 
Also note from (\ref{ineq1}) that for black hole masses much greater
than the Planck mass, $r_+$ is huge and the range of $|\L|$ 
corresponding to $C>0$ is given by the expression:
\be
|\L| > \f{(d-2)(d-3)}{2 r_+^2}~\approx 0~~. 
\la{range1}
\ee
This means that the introduction of a very small negative 
cosmological constant is sufficient to make the present formalism applicable.  

\subsection{Reissner-Nordstr\"om Black Hole}

Finally, for a $d$-dimensional Reissner-Nordstr\"om black hole with metric
\cite{mp}
\ba
ds^2 &=& - \le(  1 - \f{16\p M}{(d-2) \O_{d-2} r^{d-3}}  
+ \f{16\p Q^2}{(d-2)(d-3) r^{2(d-3)}} \ri) dt^2 \nn \\
&+& \le(  1 - \f{16\p M}{(d-2) \O_{d-2} r^{d-3}}  
+ \f{16\p Q^2}{(d-2)(d-3) r^{2(d-3)}} \ri)^{-1} dr^2
+ r^2 d\O_{d-2}^2~~,
\ea
the entropy and temperature are given by:
\ba
S_0 &=& \f{\O_{d-2}~r_+^{d-2} }{4 G_d}  \\
T_H &=&  \f{(d-3)}{2\p~r_+^{d-2} }~
\sqrt{ \le( \f{8\p G_d M}{(d-2)\O_{d-2}} \ri)^2 - 
\f{2G_dQ^2}{(d-2)(d-3)}  }  \la{sc5}
\ea
where $Q$ is the electric charge. 
Using the formula for horizon radius:
\be
r_+^{d-3} = \f{8\p G_d M}{(d-2) \O_{d-2}} +
\sqrt{ \le( \f{8\p G_d M}{(d-2) \O_{d-2}} \ri)^2  - \f{2G_d Q^2}{(d-2)(d-3)} }
~~, 
\ee
the specific heat is :
\be
C =  \le( \f{d-2}{d-3} \ri) S_0~F_2(M,Q) \la{rncq} 
\ee
and the temperature far away from extremality ($r_+ >> r_-$): 
\be
T_H = \le[ \f{d-3}{4\p} \le( \f{\O_{d-2}}{4G_d} \ri)^{1/(d-2)} \ri]~
S_0^{-1/(d-2)} ~F_1(M,Q) \la{rntemp} 
\ee
where
\ba
F_1 (M,Q) &=& 1 - 2G_d \le(  \f{d-2}{d-3} \f{\O_{d-2}Q}{16\p G_d M} \ri)^2 
\nn \\
F_2 (M,Q) &=& \f{\sqrt{  \le(\f{8\p G_d M}{(d-2) \O_{d-2}}\ri)^2 
- \f{ 2G_d Q^2}{ (d-2)(d-3)}} }
{ \f{8\p G_dM}{(d-2)\O_{d-2}} - \f{(d-2)}{(d-3)}
\sqrt{  
\le( \f{8\p G_d M}{(d-2) \O_{d-2} }\ri)^2 - \f{2G_d Q^2}{(d-2)(d-3)} }}~~.  
\nn
\ea
It can be shown that in the range: 
\be
\le[ \f{2(d-2) \O_d^2}{(8\p)^2(d-3) G_d}  \ri] Q^2 <   M^2 
< \le[ \f{2(d-2)^3 \O_d^2}{(8\p)^2(2d-5) G_d}  \ri] Q^2~~, 
\ee
$C >0$, i.e. the black hole is thermodynamically stable, 
and the leading order entropy correction from (\ref{corr3}) is: 
\be
{\cal S} = S_0 - \f{(d-4)}{2(d-2)} \ln S_0  + \cdots  ~~. 
\la{rnent}
\ee
It is interesting to note that the fluctuation corrections in this case
vanish for $d=4$. 

Since the string theoretic black holes in $5$ and in $4$ dimensions, whose 
entropy was reproduced by $D$-brane state counting, fall into the 
class of Reissner-Nordstr\"om black holes, we do not consider them
separately. 

\section{Entropy Formula from Microscopics}

We saw in the above, that to calculate the entropy corrections,
an exact entropy formula $S(\b)$ was not required. 
However, now we would like to investigate as to what kind
of functions would be sufficient to reproduce the above correction, as that
might shed some light on the nature of microscopic theory governing black
hole dynamics. 

Let us return to  Eq. (\ref{corr1}), which as mentioned
before, predicts leading correction to the equilibrium entropy in terms
of $S_0''$, the second derivative of the {\it exact} entropy function
$S(\b)$, evaluated at the equilibrium temperature $\b_0$. As stated
before, if we wish to
specialize to black holes, we must first make the following identifications:
\ba
\b_0 &=& \f{1}{T_H} ~~, \\
S_0 = S(\b_0)  &=& \f{A_H}{4G_d}  ~~,
\ea
where the first relation follows from the assumption that the black hole is
at equilibrium at Hawking temperature. 
It is clear that this set of information is inadequate to determine
the functional form of $S_0''$, which is independent of both
$S_0$ and $\b_0$.
Geometrically, the quantities $\b_0$ and $S_0$ represent the abscissa
and ordinate of the extremum of the parabola which approximates
the actual curve near extremum. The slope $S_0''$ on the other hand is
still an independent quantity.

To determine $S_0''$, we may assume any specific form
of the function $S(\b)$, which can be well approximated by the
parabolic form (\ref{ent1}). 
The only stipulation is that such a form must admit of an extremum 
at some value $\b_0$ of $\b$. 
If we assume an underlying CFT, then from modular invariance of the partition 
functions of $(1+1)$-dimensional
conformal field theories it follows that $S(\b)$ 
is of the following form (see argument of
exponential function in Eq(2.7) in \cite{carlip} ):
\be
S(\b) = a\b + \f{b}{\b}
\la{cft1}
\ee
where $a,b$ are constants.  However, we will choose a more general form:
\be
S(\b) = a\b^m + \f{b}{\b^n}
\la{ent2}
\ee
where $m,n,a/b > 0$. 
We will show later that (\ref{ent2}) is also consistent with 
assumptions made in the context of Euclidean black hole thermodynamics
\cite{euclidean}. The above expression has an extremum at 
\be
\b_0 =  \le( \f{nb}{ma} \ri)^{{1}/(m+n)}  \equiv  \f{1}{T_H}~~~.
\la{temp1}
\ee
Expanding around $\b_0$, by computing the second derivative, we get:
\be
S(\b) = \a~\le(a^n b^m \ri)^{1/(m+n)} + \f{1}{2}~\c
\le( a^{n+2}b^{m-2} \ri)^{{1}/(m+n)}~(\b - \b_0)^2 + \cdots ~,
\ee
where    
\ba
 \a &=& (n/m)^{m/(m+n)} + (m/n)^{n/(m+n)}  \nn \\
\c &=& (m+n) m^{(n+2)/(m+n)} n^{(m-2)/(m+n)} \nn
\ea
are constants.
Comparing with (\ref{ent1}), we find: 
\ba
S_0 &=& \a~\le(a^n b^m \ri)^{{1}/(m+n)} ~~,\\
S_0'' &=& \c~\le( a^{n+2}b^{m-2} \ri)^{1/(m+n)}~~.
\ea
Next, we invert these equations to get $a,b$ in terms of $S_0$ and
$S_0''$ :
\ba
a &=& \f{\a^{(m-2)/2}}{\c^{m/2}}~\le( S_0''\ri)^{m/2}
S_0^{-(m-2)/2} \\
b &=& \le[ \f{\a^{(n+2)/2}}{\c}  \ri]^{-n/2}
S_0^{(n+2)/2} \le( S_0'' \ri)^{-n/2}~~~.
\ea
These values of $a$ and $b$ can now be substituted in
(\ref{temp1}) to obtain:
\be
\b_0 = \f{1}{T_H} = \le( \f{n}{m} \ri)^{1/(m+n)} \sqrt{\f{S_0}{S_0''}~
\f{\c}{\a}}~~~.
\ee
The last relation can be inverted to 
express $S_0''$ in terms of $S_0$ and $T_H$ :
\be
S_0'' = \le[ \le(\f{\c}{\a}\ri)~
\le(\f{n}{m}\ri)^{2/(m+n)} \ri] S_0 T_H^2~.
\ee
The above relation completely specifies
$S_0''$ (and hence entropy corrections) in terms of Bekenstein-Hawking entropy
and Hawking temperature, and is a direct consequence of our
assumed form of exact entropy function (\ref{ent2}).
Again, the factors in square brackets are independent of black hole
parameters and can be dropped.  
Substituting in (\ref{corr1}), we get:
\be
{\cal S} = \ln \r = S_0 - \f{1}{2} \ln \le( S_0 T_H^2 \ri) + \mbox{smaller terms}
\la{corr2}
\ee
Note that the leading order 
correction is independent of $m,n$ and the latter has an effect only 
in the sub-leading terms.

Let us examine the implications for various black holes.  
{}From (\ref{temp4}), we see that $S_0 = C$ for BTZ. Thus, the 
correction (\ref{corr2}) becomes identical to (\ref{corr3}) ! 
Thus, the general 
form of the function $S(\b)$ in (\ref{ent2}) is
sufficient to arrive at this correction, and not just the specific
$m=1=n$ case as dictated from CFT. This makes the correction quite robust.

Again for Schwarzschild-anti-de Sitter black holes, in the positive 
specific heat domain (\ref{ineq1}), it follows from 
(\ref{adscv1}) that $C = (d-2) S_0$ to leading
order. Then, from (\ref{corr2}), we get:
\be
{\cal S} = \ln \r = S_0 - \f{1}{2} \ln \le(C T_H^2 \ri) +  \cdots
\ee
which agrees with the fluctuation correction formula (\ref{corr3}).
Hence, the logarithmic correction term will have identical coefficient as
in (\ref{scadscorr2}). 
The generalization to charged black holes is also straightforward. 

It may be noted that to demonstrate the 
instability of Schwarzschild black holes, it was assumed in 
\cite{euclidean} that the exact entropy function $S(\b)$ is 
obtained from the equilibrium formula (i.e. the 
area law) $S(\b_0)$, by simply substituting $\b_0 \rightarrow \b$.
While this is sufficient for the area law to be valid at Hawking 
temperature, it is certainly not necessary. However, we show here
that the above assumption is consistent with the entropy 
function (\ref{ent2}) that we chose in the last section. 

Again consider BTZ black holes. Using (\ref{temp3}) and $\b_0 = 1/T_H$,
we get at equilibrium :
\be
S (\b_0 ) = \f{\p^2\ell^2}{G_3} \f{1}{\b_0}~~.
\ee
Following \cite{euclidean} if we assume that the above continues to hold
for any $\b$, then we get :
\be
S (\b ) = \f{\p^2\ell^2}{G_3} \f{1}{\b}~~.
\la{btzapprox5}
\ee
Also from (\ref{temp3}) and the expression 
$r_+ = \sqrt{8G_3M}\ell$, we can write: 
\be
\b_0 = \f{ \p\ell}{ \sqrt{2G_3 M}}~.
\ee
This implies that for BTZ black hole mass much greater than the Planck mass,
$1/\b_0 \gg \b_0$. Also for $M/M_{Pl} \rightarrow \infty$, let us assume that
the above inequality holds for temperatures slightly away from equilibrium 
(since for truly macroscopic black holes, fluctuations are expected to be
small). Then, from Eq.(\ref{ent2}) it follows that :
\be
\lim_{M/M_{Pl} \rightarrow \infty} S(\b) = \f{b}{\b^n} ~~.
\la{btzapprox6}
\ee
Comparing (\ref{btzapprox5}) and (\ref{btzapprox6}) we find:
\ba
n&=&1 \nn \\
\& ~~~~ b &=& \f{(\p \ell)^2}{G_3} ~~.\nn
\ea
In other words, the assumed relation (\ref{btzapprox6}) is perfectly consistent
with the our general chosen form (\ref{ent2}) in the limit of macroscopic
black holes. This can also be seen from the fact that if (\ref{btzapprox6})
holds with $n=1$, 
then it implies $S_0'' \sim \b^{-3} \sim S_0^3 $, which using (\ref{corr1}) 
in turn implies the required corrected entropy 
$$ {\cal S} = S_0 - \f{3}{2} \ln S_0 + \cdots $$

Similar conclusions follow for AdS-Schwarzschild black hole in $d$-dimensions.
For details, see \cite{dmb}. 

\section{Conclusions}

To summarise, we have shown here that small fluctuations
around equilibrium give rise to logarithmic corrections to the 
thermodynamic entropy of a system. When applied to black holes,
they are of the form $-k \ln(A)$. For BTZ black holes, we showed
that $k=3/2$, as found earlier in several other approaches. We also
calculated $k$ for AdS-Schwarzschild and Reissner-Nordstr\"om black
holes for all spacetime dimensions. Finally, we explored certain exact
entropy functions, which may naturally arise in microscopic theories of black holes,
and which give rise to identical entropy corrections. 

It is interesting to note that under certain assumptions,
when black holes are thought to be composed of mutually non-interacting
but distinguishable 
particles in the {\it grand canonical ensemble} with fluctuating 
particle number, the entropy of the resultant system turns out to be 
proportional to area, followed by log area term \cite{ms},
although in the latter case the log correction 
occurs with a positive sign. 
Also, it might be useful to understand 
the implications of our corrections for the holographic hypothesis.
We hope to report on it in the future. 

\vs{.4cm}
\no
{\bf Acknowledgements}

The author would like to thank P. Majumdar and R. K. Bhaduri for 
numerous fruitful discussions, correspondence and for
collaboration, on which paper \cite{dmb} was based;
G. Kunstatter for useful discussions; S. Carlip and V. Husain for useful
correspondence which included several interesting comments and suggestions;
A. Dasgupta for bringing to his attention refs. \cite{mann} 
and J. Bland for comments. This work was supported by 
the Natural Sciences and Engineering Research Council of Canada.

\end{document}